\documentclass[nobibnotes,superscriptaddress,showpacs,twocolumn]{revtex4}

\usepackage{amsmath,amsfonts,bm}
\usepackage{graphicx}

\begin{document}
\title{Anomaly-Free Constraints in Neutrino Seesaw Models}
\date{\today}

\author{ D.~Emmanuel-Costa}
\email{david.costa@ist.utl.pt}
\affiliation{Departamento de F\'{\i}sica and Centro de F\'{\i}sica Te\'orica de
Part\'{\i}culas (CFTP),
Instituto Superior T\'ecnico, Avenida Rovisco Pais, 1049-001 Lisboa, Portugal}
\author{Edison T. Franco}
\email{edisontf@ift.unesp.br}
\affiliation{Instituto de F\'\i sica Te\'orica, Universidade Estadual Paulista, Rua Doutor Bento Teobaldo Ferraz 271 - Bloco II, 01140-070 - S\~ao Paulo, SP, Brazil}
\affiliation{Departamento de F\'{\i}sica and Centro de F\'{\i}sica Te\'orica de
Part\'{\i}culas (CFTP),
Instituto Superior T\'ecnico, Avenida Rovisco Pais, 1049-001 Lisboa, Portugal}
\author{R. Gonz\'{a}lez Felipe}
\email{gonzalez@cftp.ist.utl.pt}
\affiliation{Area Cient\'{\i}fica de F\'{\i}sica, Instituto Superior de Engenharia
de Lisboa, Rua Conselheiro Em\'{\i}dio Navarro 1, 1959-007 Lisboa, Portugal}
\affiliation{Departamento de F\'{\i}sica and Centro de F\'{\i}sica Te\'orica de
Part\'{\i}culas (CFTP),
Instituto Superior T\'ecnico, Avenida Rovisco Pais, 1049-001 Lisboa, Portugal}

\begin{abstract}
The implementation of seesaw mechanisms to give mass to neutrinos in the presence of an anomaly-free $U(1)_X$ gauge symmetry is discussed in the context of minimal extensions of the standard model. It is shown that type-I and type-III seesaw mechanisms cannot be simultaneously implemented with an anomaly-free local $U(1)_X$, unless the symmetry is a replica of the well-known hypercharge. For combined type-I/II or type-III/II seesaw models it is always possible to find nontrivial anomaly-free charge assignments, which are however tightly constrained, if the new neutral gauge boson is kinematically accessible at LHC. The discovery of the latter and the measurement of its decays into third-generation quarks, as well as its mixing with the standard $Z$ boson, would allow one to discriminate among different seesaw realizations.
\end{abstract}
\vspace*{2mm}
\maketitle

\section{Introduction}
The smallness of neutrino masses is elegantly explained in the context of seesaw models. Since in the standard model (SM) neutrinos are massless, new physics is required beyond the model to account for their tiny masses. The simplest possibility consists of the addition of singlet right-handed neutrinos, which seems natural once all the other fermions in the SM have their right-handed counterparts. In this framework, known as a type-I seesaw mechanism~\cite{seesaw:I}, neutrinos acquire Majorana masses at tree level via a unique dimension-five effective operator. Alternatively, neutrino masses can be generated by a type-II seesaw~\cite{seesaw:II} through the tree-level exchange of $SU(2)$-triplet scalars, which are color-singlets and carry unit hypercharge. A third possibility, often referred as type-III seesaw~\cite{Foot:1988aq}, consists of the introduction of color- singlet $SU(2)$-triplet fermions with zero hypercharge.

Adding to the theory new fermions with chiral couplings to the gauge fields unavoidably raises the question of anomalies, i.e. the breaking of gauge symmetries of the classical theory at the quantum level. For consistency, anomalies should be exactly canceled. In the SM this cancellation occurs for each fermion generation with the hypercharge assignment. On the other hand, in many SM extensions and, particularly, in various grand unified and string models, extra $U(1)$ factors naturally appear, thus bringing additional anomaly cancellation constraints~\cite{Langacker:2008yv}. These extra factors also typically lead to a richer phenomenology that might have definite signatures at the forthcoming Large Hadron Collider (LHC) experiments.

In this paper we address the above question in the context of minimal extensions of the SM, based on the gauge structure $SU(3)_{C}\otimes SU(2)_{L} \otimes U(1)_{Y} \otimes U(1)_X$ and coexisting with a seesaw-type mechanism for neutrino masses. In this framework, the heavy (singlet or triplet) fermions responsible for the seesaw will acquire large Majorana masses, after the new gauge symmetry is spontaneously broken by singlet scalar fields with nontrivial $U(1)_X$ charges. As it turns out, the axial-vector~\cite{Adler:1969gk} and mixed gravitational-gauge~\cite{Delbourgo:1972xb} anomaly-free conditions are quite constraining. Furthermore, Yukawa interactions, necessary to give masses to SM fermions as well as to implement the seesaw mechanism, restrict the number of Higgs doublets required by the theory.

\section{Anomaly constraints on seesaw}
We consider a simple extension of the SM with minimal extra matter content, so that neutrinos obtain seesaw masses and all fields are nontrivially charged under a new $U(1)_X$ gauge symmetry. We include singlet right-handed neutrinos $\nu_R$ and color-singlet $SU(2)$-triplet fermions $T$ with zero hypercharge to implement type-I and type-III seesaw mechanisms, respectively. To give masses to quarks and leptons, 4 Higgs doublets ($H_u\,,H_d\,,H_\nu\,,H_e$) are in general required, while another 2 scalar doublets, $H_\delta$ and $H_T$, and one $SU(2)$-triplet scalar $\Delta$ are necessary to generate Yukawa terms for type-II and type-III seesaw mechanisms. Finally, a singlet scalar field $\phi$ is introduced to give Majorana masses to $\nu_R$ and $T$.

To render the theory free of anomalies the following set of constraints must be satisfied:
\begin{subequations}
\label{an}
\begin{align}
A_1=&N_{g}\left(2\,x_{q}-x_{u}-x_{d}\right) =0\,, \label{eq:A1}\\ A_2=&N_{g}\left(x_{\ell}/2+3\,x_{q}/2\right)-2\,N_T\, x_T =0\,, \label{eq:A2} \\ A_3=&N_{g}\left(x_{q}/6+x_{\ell}/2 -4\,x_{u}/3 -x_{d}/3 - x_{e}\right)=0\,, \label{eq:A3}\\
A_4=&N_{g}\left(x_{q}^{2}- x_{\ell}^{2} - 2x_{u}^{2} + x_{d}^{2} \,+\,
x_{e}^{2}\right) =0\,,
\label{eq:A4}\\
A_5=&N_{g}\left(6\,x_{q}^3+2\,x_{\ell}^3-3\,x_{u}^3-3\,x_{d}^3-x_{e}^3
\right)\nonumber\\
&-N_R\,x_{\nu}^3-3\,N_T\,x_{T}^3=0\,,
\label{eq:A5}\\
A_6=&N_{g}\left(6\,x_{q}+2\,x_{\ell}-3\,x_{u}-3\,x_{d}-x_{e} \right)\nonumber\\
& -N_R\,x_{\nu}-3\,N_T\, x_{T}=0\,, \label{eq:A6} \end{align} \end{subequations}
where $x_i$ denote the fermion charges under $U(1)_X$ (the subscript $i$ refers to the field), $N_g$ is the number of generations, $N_R$ is the number of right-handed neutrinos and $N_T$ is the number of fermionic triplets. Equations~(\ref{eq:A1})-(\ref{eq:A5}) arise from the requirement of the cancellation of the axial-vector anomaly, while Eq.~(\ref{eq:A6}) results from the cancellation of the gravitational-gauge anomaly.

Making use of Eqs.~(\ref{eq:A1})-(\ref{eq:A3}) and (\ref{eq:A6}) we can express the charges $x_{q}\,$, $x_{\ell}\,$, $x_{u}$, and $x_{d}$ in terms of $x_{e}\,$, $x_{\nu}$, and $x_{T}$ as
\begin{equation}
\label{eq:sol1236}
\begin{aligned}
x_{q}&=-\frac{N_g\,x_{e}+N_R\,x_{\nu}-5\,N_T\,x_{T}}{6\,N_g}\,,\\
x_{\ell}&=\frac{N_g\,x_{e}+N_R\,x_{\nu}+3\,N_T\,x_{T}}{2\,N_g}\,,\\
x_{u}&=-\frac{2\,N_g\,x_{e}-N_R\,x_{\nu}-N_T\,x_{T}}{3\,N_g}\,,\\
x_{d}&=\frac{N_g\,x_{e}-2\,N_R\,x_{\nu}+4\,N_T\,x_{T}}{3\,N_g}\,.
\end{aligned}
\end{equation}
Substituting these expressions into Eq.~(\ref{eq:A4}) for the $A_4$ anomaly, we derive the condition
\begin{equation}
\label{eq:A4an}
A_4=-\frac{4\,N_R\,N_T\,x_{\nu}\,x_{T}}{N_g}=0\,,
\end{equation}
which, clearly, has only the trivial solutions $N_R=0$, $N_T=0$, $x_{\nu}=0$, or $x_{T}=0$. Thus, we conclude that it is not possible to have an anomaly-free local $U(1)_{X}$ and, simultaneously, type-I ($N_R \neq 0$) and type-III ($N_T \neq 0$) seesaw models, unless $U(1)_{X}$ is proportional to the hypercharge $U(1)_{Y}$, in which case no new gauge symmetry is obtained. Of course, both types of seesaw could coexist if new extra matter content is added to the theory~\cite{Ma:1998dn}. In such a case, at least two extra singlets charged under $U(1)_{X}$ are needed to cancel the $A_4$ and $A_5$ anomalies.

Once $A_4=0$ is enforced, the anomaly condition $A_5$ given in Eq.~(\ref{eq:A5}) can be written in terms of $x_{\nu}$ and $x_{T}$ as
\begin{equation}
\label{eq:A5an1}
A_{5}=\frac{1}{N_{g}^{2}}\left[ N_{R}(
N_{R}^{2}-N_{g}^{2})\,x_{\nu }^{3} +3N_{T}( N_{T}^{2}-N_{g}^{2})\,x_{T}^{3}
\right]\,.
\end{equation}

\subsection{Type-I seesaw case with $N_R =N_g$}
If no fermion triplet is introduced and the type-I seesaw mechanism is responsible for neutrino masses, the anomaly constraint $A_5=0$ implies $N_R=N_g$ (the solution $x_{\nu}=0$ leads to the hypercharge). Thus, one right-handed neutrino per generation is required. In this case, all $U(1)_{X}$ charge assignments can be expressed in terms of the quark doublet charge $x_q$ and the charge ratio $\alpha \equiv -x_{e}/x_{q}\,$, as presented in Table~\ref{tab:sol}, leading to an infinite class of anomaly-free local $U(1)_{X}$ symmetries. Note that, in this parametrization, $\alpha=6$ corresponds to the hypercharge and $\alpha=3$  gives the usual $U(1)_{B-L}$ gauge symmetry.  Any other value of $\alpha$ would correspond to a new anomaly-free gauge symmetry.
\begin{table}
\caption{\label{tab:sol} $U(1)_{X}$ fermion charges (normalized with $x_q$) as a function of the ratio $\alpha \equiv -x_{e}/x_{q}$ in different minimal seesaw realizations.}
\begin{ruledtabular}
\begin{tabular}{cccc}
 & Type-I & Type-III & Type-I\\
$U(1)_X$ charge & $N_R = N_g$ & $N_T = N_g$ & $N_g=3, N_R = 2$\\
\hline
$x_{u}$   & $\alpha-2$   & $\frac15\,(2+3\,\alpha)$ & $\alpha-2$ \\
$x_{d}$   & $4-\alpha$   & $\frac15\,(8-3\,\alpha)$ & $4-\alpha$ \\
$x_{\ell}$& $-3$   & $\frac15\,(9-4\,\alpha)$ & $-3$ \\
$x_{\nu}$ & $\alpha-6$   & \emph{n.\,a.}         & $4\,(\alpha-6) $ \\
$x_{T}$   & \emph{n.\,a.} & $\frac15\,(6-\alpha)$  & \emph{n.\,a.}\\
$x_{S}$   & \emph{n.\,a.} & \emph{n.\,a.}  & $5\,(6-\alpha)$
\end{tabular}
\end{ruledtabular}
\end{table}

\subsection{Type-III seesaw case with $N_T = N_g$}
In the absence of right-hand neutrinos ($N_R=0$) and invoking a type-III seesaw to give masses to neutrinos, the anomaly constraint $A_5=0$ yields $N_T=N_g$ (the solution $x_T=0$ would lead once again to the hypercharge). Therefore, a fermion triplet per family is required to cancel all the anomalies. The corresponding $U(1)_{X}$ charge assignments are displayed in Table~\ref{tab:sol} as a function of the ratio $\alpha\,$. Although in this case the hypercharge solution can be recovered by choosing $\alpha=6$, there is no value of $\alpha$ that leads to the $U(1)_{B-L}$ gauge symmetry. The choice $\alpha=3$ and $x_q=1/3$ implies $x_{u}=11/5$, $x_{d}=x_{\ell}=-1/5$, $x_{e}=-1$, and $x_{T}=1/5$, which obviously do not correspond to the correct $(B-L)$ charges. This peculiar type of solution was first found in Ref.~\cite{Barr:1986hj} and, more recently, it has been the subject of several studies~\cite{Ma:2002pf}.

\subsection{Type-I seesaw case with $N_R \neq N_g$}
Neutrino oscillation data do not demand the existence of three right-handed neutrinos to successfully implement a type-I seesaw mechanism. In fact, the presence of just two $\nu_R$ fields can account for the large leptonic mixing and the solar and atmospheric mass-squared differences. This is a predictive scenario since the lightest neutrino is massless, while the other two neutrino masses are fixed by the observed solar and atmospheric mass-squared differences. It is therefore legitimate to ask ourselves whether such a theory admits an anomaly-free $U(1)_{X}$ gauge symmetry. This would require one to relax the condition $N_R=N_g\,$ imposed by the $A_5$ anomaly constraint, which is only possible if new matter content is added to the theory. The simplest possibility is to assume a ``sterile" right-handed singlet $\nu_S$, which does not participate in the seesaw mechanism and has a $U(1)_{X}$ charge $x_S$ different from the charge $x_\nu$ of the ``active" $\nu_R$ neutrinos. Clearly, an additional scalar singlet field, $\phi_{S}$, would be necessary to generate mass for the sterile right-handed neutrino. In this framework, the anomaly conditions given in Eqs.~(\ref{eq:A5}) and (\ref{eq:A6}) must be modified. Assuming three generations, two active and one sterile right-handed neutrinos, the anomaly-free system of equations implies the charge assignments given in the last column of Table~\ref{tab:sol}. For $\alpha=6$, the hypercharge is reproduced, while $\alpha=3$ does not lead to $B-L$ since $x_{\ell,e}=-1$, $x_{u,d,q}=1/3$, $x_\nu=-4$, and $x_S=5$~\cite{Montero:2007cd}.

\section{Quark and lepton masses}
We now consider the constraints coming from the Yukawa terms responsible for the quark and lepton masses. The most general Lagrangian terms to generate fermion masses, and which are compatible with type-I, type-II, and type-III seesaw models for Majorana neutrinos, are given by
\begin{equation}
\begin{split}
&Y_{u}\,\bar{q}_L u_{R}H_{u}+Y_{d}\,\bar{q}_L
d_{R}H_{d}+ Y_{e}\,\bar{\ell}_L e_{R} H_{e}
+Y_{\nu}\,\bar{\ell}_{L}\nu_{R}H_{\nu}\\
&+Y_{T}\,\bar{\ell}_{L}i\tau_{2}T H_{T}\,+
Y_{\text{I}}\,\nu_{R}^{T}C\nu_{R} \phi
+Y_{\text{II}}\,\ell_{L}^{T}Ci\tau_{2}\ell_{L}\Delta\\
&+Y_{\text{III}}\,\mathrm{Tr } \left(T^{T}CT\right) \phi +\lambda_{\Delta}
H_{\delta} H_{\delta} \Delta+\text{H.c.}.
\label{Ly}
\end{split}
\end{equation}
The following relations then hold for the $z_i$ boson charges under $U(1)_X$: $x_{q}=x_{u}+z_u = x_{d}+z_d,\,x_{\ell}=x_{e}+z_{e},\,x_{\ell} = x_{\nu }+z_{\nu },\,x_{\ell}=x_{T}+z_{T}, \, 2x_{\nu }+z_{\phi }=0,\, 2x_{\ell}+z_{\Delta}=0, \, 2x_{T}+z_{\phi}=0$, and $2z_{\delta }+z_{\Delta}=0$.
The solutions of these equations are presented in Table~\ref{tab:yuk}, for the three minimal seesaw models considered above. We conclude that, for a type-I seesaw scenario with a right-handed neutrino per family, only one Higgs doublet $H_{u}$ is required to give masses to all fermions. On the other hand, for a minimal type-I seesaw solution with $N_{g}=3$ and $N_{R}=2$, while the same Higgs doublet $H_{u}$ is sufficient to give mass to the quarks and charged leptons, a second one, $H_{\nu}\,$, is needed to implement the type-I seesaw and give mass to light neutrinos. If a type-II seesaw is also present, then we must introduce an extra Higgs doublet $H_{\delta}\,$. Of course, for the standard hypercharge solution $(\alpha=6)$ all three Higgs doublets coincide,  $H_{u}=H_{\nu}=H_{\delta}\,$.

For the type-III seesaw case, one can see from the Table~\ref{tab:yuk} that a single Higgs doublet $H_{u}$ can simultaneously give masses to the quarks and to the light neutrinos through the seesaw. Yet, another Higgs doublet $H_{e}$ must account for the charged lepton masses. When the type-II seesaw is also introduced, one more Higgs doublet $H_{\delta}$ is needed.  Once more, for the standard hypercharge solution all three Higgs doublets can be identified as a unique doublet.

We remark that the analysis presented above should be taken just as a generic statement on the minimal number of Higgs doublets required to give mass to fermions, while keeping the theory free of anomalies. This, of course, does not preclude the presence of different Higgs doublets charged under the $U(1)_X$ symmetry.
\begin{table}
\caption{\label{tab:yuk}  $U(1)_{X}$ boson charges (normalized with $x_q$) as a
function of the ratio $\alpha \equiv -x_{e}/x_{q}$ in different minimal seesaw
realizations.}
\begin{ruledtabular}
\begin{tabular}{cccc}
& Type-I/II & Type-III/II & Type-I/II\\
$U(1)_X$ charge & $N_R = N_g$ & $N_T = N_g$ & $N_g=3, N_R = 2$\\ \hline
$z_{u}$     & $3-\alpha$   & $\frac35\,(1-\alpha)$ & $3-\alpha$ \\
$z_{d}$     & $\alpha-3$   & $\frac35\,(\alpha-1)$ & $\alpha-3$ \\
$z_{e}$     & $\alpha-3$   & $\frac15\,(9+\alpha)$ & $\alpha-3$ \\
$z_{\nu}$   & $3-\alpha$   & \emph{n.\,a.}         & $21-4\,\alpha$ \\
$z_{T}$     & \emph{n.\,a.} & $\frac35\,(1-\alpha)$ & \emph{n.\,a.} \\
$z_{\phi}$  & $2\,(6-\alpha)$ & $\frac25\,(\alpha-6)$ & $8\,(6-\alpha)$\\
$z_{\Delta}$& 6 & $\frac25\,(4\,\alpha-9)$ & 6\\
$z_{\delta}$& $-3$ & $\frac15\,(9-4\,\alpha)$  & $-3$\\
$z_{S}$  & \emph{n.\,a.} & \emph{n.\,a.} & $ 10\,(\alpha-6)$
\end{tabular}
\end{ruledtabular}
\end{table}

\begin{figure}
\includegraphics[width=8cm]{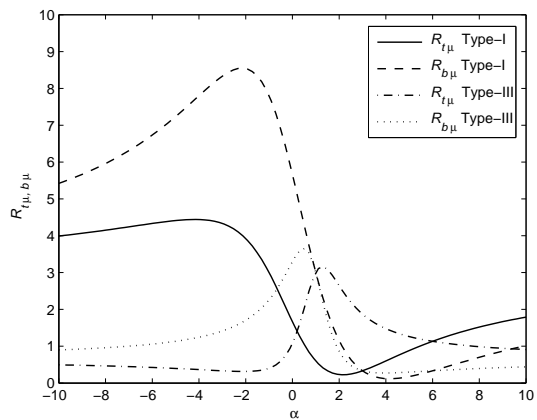}
\caption{\label{fig1} Branching ratios of the $X$ decays into quarks and muons as a function of the charge ratio~$\alpha$.}
\end{figure}
\begin{figure}
\includegraphics[width=8cm]{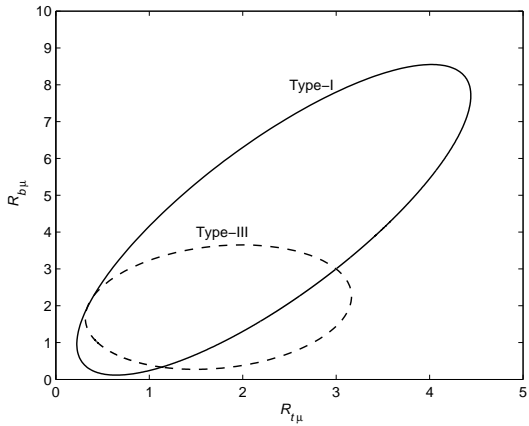}
\caption{\label{fig2} $R_{t\mu} - R_{b\mu}$ plane for type-I and -III seesaw realizations.}
\end{figure}

\section{Phenomenological implications}
One of the attractive features of theories with extra $U(1)$ gauge symmetries is their richer phenomenology, when compared with the SM. Indeed, the spontaneous breaking of the symmetry leads to new massive neutral gauge bosons $X$ (often called $Z^\prime$) which, if kinematically accessible, could be detectable at LHC. In the latter case, its decays into third-generation quarks, $pp \rightarrow X \rightarrow b\,\bar{b}$ and $pp \rightarrow X \rightarrow t\,\bar{t}$ could be used to discriminate between different models~\cite{Godfrey:2008vf}. In particular, the branching ratios of quark to $\mu^+ \mu^-$ production,
\begin{subequations}\label{branchratios}
\begin{align}
    R_{b\mu}&=\frac{\sigma(pp \rightarrow X \rightarrow b\,\bar{b})}{\sigma(pp
\rightarrow X \rightarrow \mu^+ \mu^-)}\simeq 3 K_q
\frac{x_q^2+x_d^2}{x_\ell^2+x_e^2}\,,\\
    R_{t\mu}&=\frac{\sigma(pp \rightarrow X \rightarrow t\,\bar{t})}{\sigma(pp
\rightarrow X \rightarrow \mu^+ \mu^-)}\simeq 3 K_q
\frac{x_q^2+x_u^2}{x_\ell^2+x_e^2}\,,
\end{align}
\end{subequations}
have the advantage of reducing the theoretical uncertainties~\cite{Godfrey:2008vf}. Here $K_q \sim \mathcal{O}(1)$ is a constant which depends on QCD and electroweak correction factors. Notice that these ratios depend on the quark and charged lepton $U(1)_X$ charges. Therefore, an analysis of the $R_{t\mu} - R_{b\mu}$ parameter space would allow one to determine these charges and, in turn, to distinguish the models.
Using the charge values given in Table~\ref{tab:sol} we obtain
\begin{equation}
R_{b\mu}\simeq \frac{3(17-8\alpha+\alpha^2)}{9+\alpha^2}\,,\, R_{t\mu}
\simeq \frac{3(5-4\alpha+\alpha^2)}{9+\alpha^2}\,,
\end{equation}
for the type-I seesaw cases, and
\begin{equation}
R_{b\mu}\simeq \frac{3(89-48\alpha+9\alpha^2)}{81-72\alpha+41\alpha^2}\,,\, R_{t\mu} \simeq
\frac{3(29+12\alpha+9\alpha^2)}{81-72\alpha+41\alpha^2}\,,
\end{equation}
for the type-III seesaw case. As can be seen from Figs.~\ref{fig1} and \ref{fig2}, the above branching ratios indeed allow one to discriminate among the models.

\begin{figure}
\includegraphics[width=8cm]{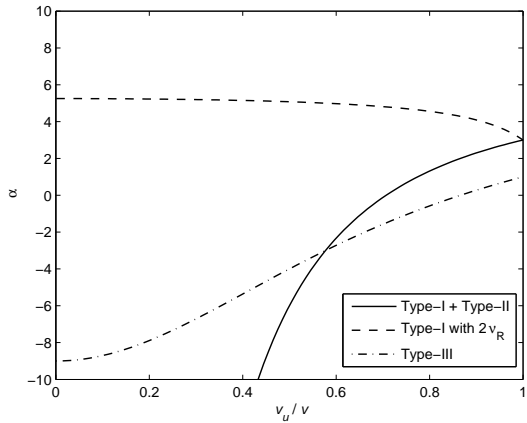}
\caption{\label{fig3} The $U(1)_X$ charge ratio $\alpha$ as a function of the ratio $r_u = v_u/v$ for vanishing $Z-X$ mixing.}
\end{figure}

Electroweak precision data also severely constrain any mixing with the ordinary $Z$ boson to the subpercent level~\cite{Langacker:2008yv}. In the models under consideration, the $Z-X$ mixing appears due to the presence of Higgs bosons which transforms under the SM gauge group and the new $U(1)_X$ Abelian gauge symmetry. Generically, such mixing is proportional to the combination~$g_Z g_X\sum_i y_ix_iv_i^2\,$, where $g_i$ are the gauge couplings, and $v_i\equiv\langle H_i\rangle$ are the vacuum expectation values of the Higgs doublets with $\sum_i v_i^2=v^2$. Hence, the requirement of vanishing $Z-X$ mixing imposes an additional constraint on the $U(1)_X$ charges. In particular, for minimal setups with only two Higgs doublets, it is possible to determine $\alpha$ as a function of a single ratio, e.g. $r_u = v_u/v$. For type-I seesaw with $N_R=N_g$ together with type-II seesaw, we find
\begin{equation}
 \alpha=6-\frac{3}{r_u^2}\,,
\end{equation}
while a type-I seesaw with only 2 right-handed neutrinos yields
\begin{equation}
 \alpha=\frac{21-18r_u^2}{4-3r_u^2}\,.
\end{equation}
Finally, a type-III seesaw realization implies
\begin{equation}
 \alpha=\frac{12r_u^2-9}{1+2r_u^2}\,.
\end{equation}
Once again, the different models are clearly distinguishable, as can be seen from Fig.~\ref{fig3}.

\section{Conclusion}
In summary, we have implemented minimal seesaw realizations in extensions of the SM with a local $U(1)_X$ symmetry. We have shown that type-I and type-III seesaw mechanisms cannot be simultaneously realized, unless the $U(1)_X$ symmetry is a replica of the standard hypercharge or new fermionic fields are added to the theory. When combined type-I/II or type-III/II seesaw models are considered, we have seen that it is always possible to assign nontrivial anomaly-free charges to the fields.

The LHC era is an exceptional opportunity to explore and discover physics beyond the SM. The minimal seesaw models studied here are well-motivated (nonsupersymmetric) extensions of the SM. The discovery of a new neutral gauge boson with a mass up to a few TeV and typical electroweak scale couplings would be one of the first evidences of a new TeV scale sector. We have demonstrated how the gauge boson signatures at LHC, like its decays into third-generation quarks and mixing with the ordinary $Z$ boson, could allow one to constrain these seesaw models and discriminate among them. Furthermore, these indirect seesaw signatures have the advantage of being independent of the Yukawa interactions of the heavy neutrino fields with the SM charged leptons. The latter interactions are expected to be very suppressed and lead to unobservable production rates of the heavy fields, if the extra $U(1)$ gauge symmetry is broken at TeV energies. In this scenario, the seesaw mechanism could still be natural if some flavor symmetries that correctly reproduce the light neutrino masses, while keeping sizable Yukawa couplings, are invoked~\cite{Kersten:2007vk}. Clearly, there is much more phenomenology with the seesaw than the one considered here. In particular, when the decays of $X$ to heavy Majorana neutrinos are kinematically allowed, like-sign dilepton signals offer another interesting possibility for studying the physics of new gauge bosons at LHC~\cite{Keung:1983uu,delAguila:2007ua}.

\section*{Acknowledgements}
This work was partially supported by Funda\c{c}\~{a}o para a Ci\^{e}ncia e a Tecnologia (FCT, Portugal) through the Projects No. POCTI/FNU/ 44409/2002, PDCT/FP/63914/2005, PDCT/FP/63912/2005, and CFTP-FCT UNIT 777, and also partially funded by POCTI (FEDER). The work of E.T.F. was supported by Funda\c{c}\~{a}o de Amparo \`{a} Pesquisa do Estado de S\~{a}o Paulo (FAPESP) under the Project No. 03/13869-3.

\end{document}